\documentclass[sigconf]{acmart}
\AtBeginDocument{%
  }

\setcopyright{acmlicensed}
\copyrightyear{2025}
\acmYear{2025}
\acmDOI{XXXXXXX.XXXXXXX}
\acmConference[AM.ICAD 2025]{Audio Mostly \& ICAD Joint Conference}
{June 30--July 04, 2025}{Coimbra, Portugal}
\acmISBN{978-1-4503-XXXX-X/2018/06}

\begin{document}



\title[Beyond Universality]{Beyond Universality: Cultural Diversity in Music and Its Implications for Sound Design and Sonification}








\author{Rub\'en Garc\'ia-Benito}
\orcid{0000-0002-7077-308X}
\email{rgb@iaa.es}
\affiliation{%
  \institution{Instituto de Astrofísica de Andalucía (IAA), CSIC}
  \city{Granada}
  \country{Spain}  
}

\affiliation{%
  \institution{\textit{Todos los Tonos y Ayres} Ensemble}
  \city{Granada}
  \country{Spain}  
  }


\begin{abstract}

The Audio Mostly (AM) conference has long been a platform for exploring the intersection of sound, technology, and culture. Despite growing interest in sonic cultures, discussions on the role of cultural diversity in sound design and sonification remain limited. This paper investigates the implicit biases and gaps within the discourse on music and sound aesthetics, challenging the notion of music as a ``universal language''. Through a historical and cross-cultural analysis of musicology and ethnomusicology, the profound influence of cultural context on auditory perception and aesthetic appraisal is highlighted. By drawing parallels between historical music practices and contemporary sound design, the paper advocates for a more inclusive approach that recognizes the diversity of sonic traditions. Using music as a case study, we underscore broader implications for sound design and sonification, emphasizing the need to integrate cultural perspectives into auditory design practices. A reevaluation of existing frameworks in sound design and sonification is proposed, emphasizing the necessity of culturally informed practices that resonate with global audiences. Ultimately, embracing cultural diversity in sound design is suggested to lead to richer, more meaningful auditory experiences and to foster greater inclusivity within the field.

\end{abstract}



\received{XXX}
\received[revised]{YYY}
\received[accepted]{Z May 2025}

\maketitle

\section{Introduction}
\label{sec:intro}

The Audio Mostly (AM) conference has long served as a platform for exploring the intersection of sound, technology, and culture. The 2024 theme, \textit{Explorations in Sonic Cultures}, highlighted the potential of sound design and sonification as accessible tools capable of transcending technical boundaries to engage diverse audiences. While one might have expected such a theme to emphasize areas like aesthetics, auditory perception, cognitive science, and universals in music, these topics have historically played a relatively minor role in the conference discourse. Despite these topics'  limited prominence, their implicit relevance and growing importance suggest that they remain vital to the ongoing exploration of culture’s role in sound.

As the sound design and sonification community continues to grow and diversify, addressing  cultural biases and gaps in the discourse has become increasingly crucial. This paper seeks to bridge these gaps by exploring the role of cultural diversity in shaping auditory aesthetics and perception.

\begin{figure*}[t]
\centering
\includegraphics[width=1.0\linewidth]{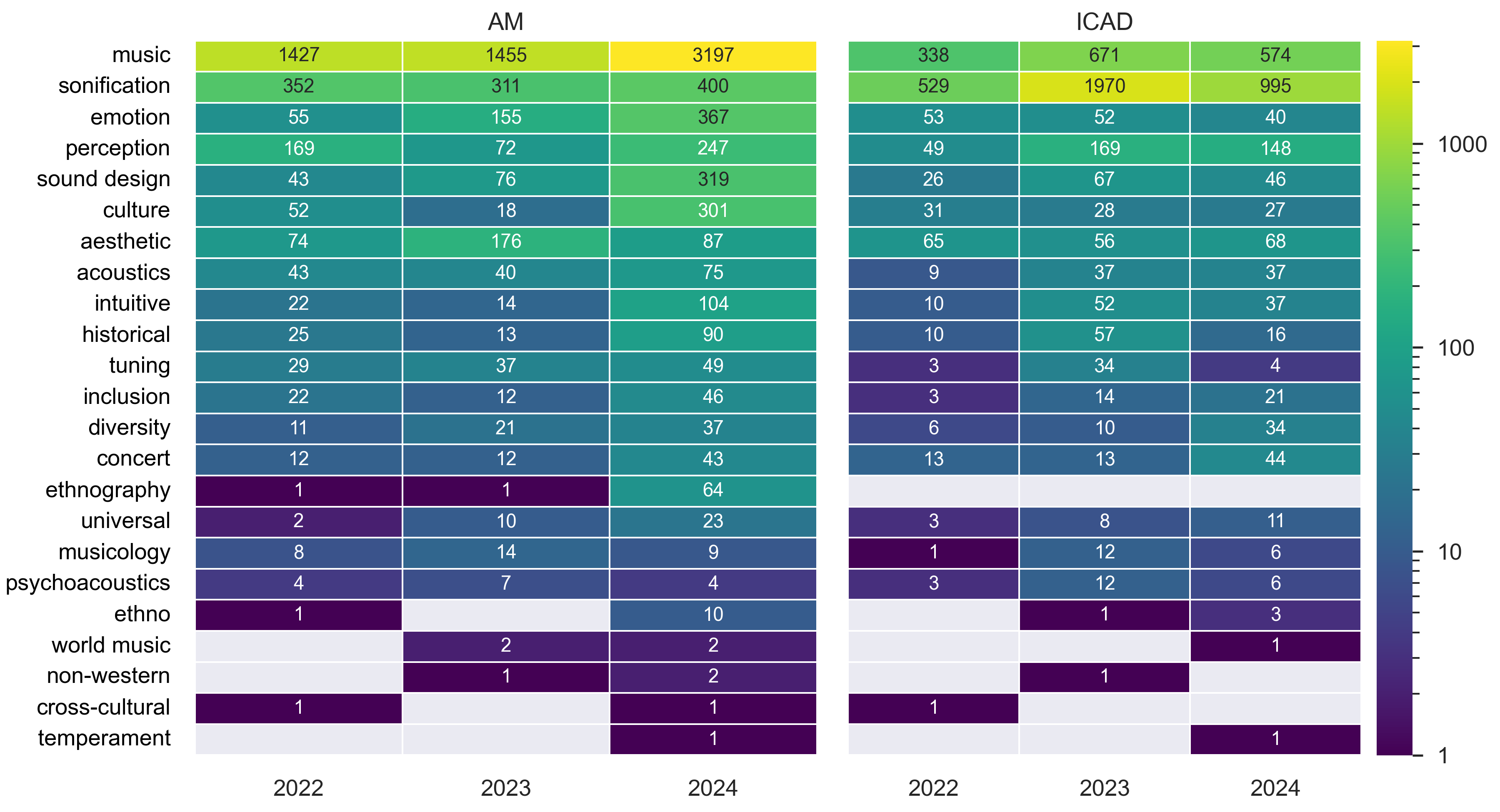}
\caption{Side-by-side heatmaps displaying the frequency of selected words in the AM and ICAD proceedings from 2022 to 2024. The vertical axis lists the chosen words, ordered by their total count in AM across the three years; this same order is applied to the ICAD heatmap to facilitate direct comparison.
The horizontal axis represents the years. Each cell is color-coded according to the word count using a logarithmic scale, with a shared normalization across both conferences to allow meaningful visual comparison. The corresponding count is displayed in each cell.}
\Description{Two heatmaps comparing word usage in AM and ICAD from 2022 to 2024. Words are ordered by total frequency in AM and shown on the vertical axis; years are on the horizontal axis. A shared color scale reflects the frequency intensity.}
\label{fig:am_kws}
\end{figure*}

To contextualize this exploration, Figure \ref{fig:am_kws} presents a heatmap visualization of selected keyword frequencies in the AM conference proceedings from 2022 to 2024. For reference and comparison, a parallel heatmap for ICAD (International Conference on Auditory Display) is shown alongside. A custom Python package was used to extract and analyze the full text of each PDF, matching pre-selected keywords to generate word counts per publication \cite{Garcia-Benito2025}. The software is available at Zenodo\footnote{\url{https://doi.org/10.5281/zenodo.15460201}} and PyPI\footnote{\url{https://pypi.org/project/pdf-word-counter/}}.
By relying on full-text analysis through automated text matching—rather than limiting the scope to abstracts or titles—this method aims to capture both explicit and implicit discussions of key themes. Although this approach prioritizes breadth over depth, it effectively reveals overarching discursive patterns. The heatmap is intended not as a comprehensive discourse analysis, but as an accessible entry point for identifying historical trends and potential gaps in cultural engagement across the AM community. 
 

The selected keywords are displayed on the vertical axis, ordered by their total frequency across the three years of AM proceedings. The same order is applied to the ICAD heatmap to enable direct comparison. Each cell in the matrix is color-coded to represent the word count, with the numerical value displayed at its center. The keywords were chosen to align with the thematic focus of this study, with some terms grouped to reflect broader conceptual categories. For instance, the category ``ethno'' aggregates the terms ``ethnic'' and ``ethnomusicology'' but explicitly excludes ``ethnography'', which is treated separately; ``cross-cultural'' includes variants such as ``cross cultural'', ``multi-cultural'', and ``multi cultural''; and ``inclusion'' combines ``inclusion'' and ``inclusivity''. Some general terms were adjusted to avoid double counting due to shared lexical stems—for example, ``music'' was adjusted by subtracting occurrences of ``ethnomusicology'' and ``musicology'', which are accounted for in other categories. After aggregation, redundant or overlapping terms were removed to prevent inflated frequencies. Technical terms such as ``sonification'' and ``sound design'' were also included to situate the cultural keywords within the broader thematic scope of each conference. This approach prioritizes interpretability and consistency over lexical granularity,  aiming to highlight broader patterns in the discourse.

The heatmap reveals several key trends. First, music has consistently been a central topic of discussion within the AM community, with its presence more than doubling in 2024 compared to the previous two years. This underscores the enduring importance of music in sound design and sonification, a trend also reflected in foundational texts such as \textit{The Sonification Handbook} \cite{Hermann2011}, which contains 727 references to music and 78 to emotion. Second, terms like ``emotion'' and ``perception'' have seen a marked increase in usage, particularly in 2024, aligning with the conference’s focus on the human experience of sound. ``Aesthetic'', another closely related term, ranks seventh in overall frequency, further emphasizing the community’s interest in the sensory and emotional dimensions of sound.

Although many papers in the AM community focus on sound design, sonification, and other technical or non-musical applications, music remains a foundational element—appearing three to eight times more frequently than these technical terms in the proceedings. This is evident in the heatmap, where ``music'' is not only the most frequently mentioned keyword but also the first in the selected list. This persistent imbalance suggests an implicit community consensus regarding music’s foundational role. When placed in relation to AM, the ICAD results offer a useful comparative point. ICAD shows a higher baseline for “sonification,” which ranks first among the selected keywords, yet ``music'' still appears in second position, indicating its continued relevance. Music serves as a shared conceptual framework for understanding sound, encapsulating cultural, emotional, and aesthetic dimensions often implicit in sound design and sonification. Building on the prominence of music within the community discourse, this paper draws on music as a shared conceptual reference to foreground its centrality, and to bridge the gap between technical sound design practices and the broader cultural, aesthetic, and emotional dimensions of sound.

While music remains a dominant theme in the discourse, the analysis also reveals significant underexplored areas—particularly regarding how cultural backgrounds influence music, aesthetics, sound design, and sonification. Although terms such as  ``culture'' and ``inclusion'' did show an increase in 2024, this is partly attributable to the AM conference theme, \textit{Explorations in Sonic Cultures}, which is referenced throughout the proceedings, rather than a substantial shift in engagement with cultural diversity. For instance,  ``culture'' often appears in generalized or thematic contexts rather than in discussions that address cross-cultural comparison or the influence of diverse cultural frameworks on sonic practices. Similarly, while ``tuning'' has slightly increased in frequency, its usage is primarily related to technical parameter adjustments—such as ``tuning parameters''  or ``fine-tuning models''—rather than references to tuning systems or musical temperaments.

In fact, terms that directly signal cross-cultural or non-Western\footnote{The terms ``Western'' and ``non-Western'' are used in this paper as a shorthand to describe broad cultural distinctions in music and aesthetics, following common usage in non-specialist discourse. While these terms are not without limitations—for example, they can obscure the diversity within and beyond the so-called ``Western'' cultural sphere—the dichotomy remains a practical shorthand for discussing broad cultural differences, particularly for non-experts. In this paper, we use these terms with the understanding that they are simplifications, and we encourage readers to consider the rich diversity within and beyond these categories. For a more nuanced discussion of these terms, see \cite{Said1978,Born2000}.} perspectives remain scarce. ``Cross-cultural'' appear only once in AM 2024, and discussions of ``ethnography'' or ``ethnographic'' primarily focus on surveys within specific communities rather than comparative or multicultural approaches. Moreover, terms such as  ``non-Western'',  ``world music'' or ``temperament'' (in the context of tuning systems), are notably absent. This suggests a need for greater engagement with the diversity of sonic aesthetics and cultural inclusivity in sound design and sonification, with particular attention to instances where the object is artistic creation, such as musification.

Over the past two decades, the sound design community has expanded rapidly, attracting participants from a wide range of disciplines. Notably, this includes fields not traditionally associated with sound, such as astronomy 
\cite{Zanella2022,Garcia-Benito2023}, 
where researchers are increasingly adopting sonification for data analysis and communication. However, as the field grows, the lack of cross-cultural perspectives remains a critical gap. While a significant body of work has focused on the practical application and evaluation of aesthetics and music, theoretical or philosophical approaches as a primary subject have been comparatively rare \cite{Vickers2006,Straebel2010,Vickers2016,Cunningham2023}. Furthermore, despite the frequent mention of culture, few studies have explicitly addressed its role in shaping sound and aesthetics \cite{Langlois2008,Jeon2015}. These topics have not been explored with the same depth or breadth as in other disciplines \cite{Pei-Luen2022,Hyde2016}, which emphasize the critical importance of cultural considerations.

By examining the role of culture in aesthetics—a frequently overlooked topic—this paper challenges existing assumptions and advocates for a more inclusive approach to sound design and sonification. Using music and aesthetics as a unifying framework, it introduces new perspectives on cultural inclusivity, addressing gaps identified in past AM conferences. These insights aim to inspire the development of more inclusive practices and evaluation methods, benefiting both newcomers and established members of the sound design and sonification community.

\section{Culture, Music, and its Universality}

Although there is a closely similar precedent in the English language \cite{lady1826}, poet Henry Wadsworth Longfellow is credited with one of the most famous sentences (included in the chapter titled ``Ancient Spanish Ballads'' from his book ``Outre-Mer'' \cite{Longfellow1835}) that permeates common musical knowledge: ``\textit{Music is the universal language of mankind},—poetry their universal pastime and delight''. Although the intended universality of Longfellow was probably more in the sense of a shared admiration for poetry and music by all humankind, the idea of music's ability to communicate and resonate across diverse cultures has nevertheless become a widely recognized statement promoting cross-cultural understanding.

The question of whether music can be considered a language requires an examination from the perspective of philosophy of language and philosophy of music \cite{Clark1982,Jackendoff2011}. The focus of this paper will be on exploring the ideas and assumptions underlying the claim of the universality of music.

An instructive analogy can be drawn from astronomy, where two complementary approaches are used to study galaxy evolution. One examines the fossil record encoded in stellar populations, using the properties of stars—formed at different times and under different conditions—to reconstruct a galaxy’s formation history. The other uses redshift surveys to observe galaxies at various stages of cosmic time, offering a direct view of their evolution.

Similar approaches can be found in other academic fields, particularly in the investigation of human 'populations' \cite{Chrisomalis2006}. On one hand, we can compare different cultures at a given moment of time and contrast their differences and similarities. On the other hand, we can also study the evolution of musical practices and aesthetics of a given culture across time. This parallel underscores the importance of both synchronic and diachronic perspectives when investigating music's perceived universality.

\subsection{Historical Musicology: Reconstructing Lost Soundscapes}
\label{sec:hm}

Music, as an ephemeral activity unfolding in the time domain, could not be captured as it was performed until the advent of recording devices in the 20th century. 
However, we have just barely a century of (limited) recorded music and thus we do not have any data of previous periods in history. 
Historical musicology studies music from a historical point of view and allows us to recover and reconstruct those missing pieces. Obviously, it has to rely on written records. Using archival documents, manuscripts, theory books, original scores and even contracts, bills or letters, musicologists are able to decode the practices of the past across time. Many disciples line together to address this daunting enterprise, from semiotic or sociological studies to historical methods, music theory, acoustics or musical practice. The analysis of period instruments, the fossil record of the instruments used at a particular time and society, help us to unlock some unsolved puzzles from the written materials and give voice to the soundscape and acoustical aesthetics of that period. Other sources like sculptures and paintings, that is, the study of iconography, help to complement performance and cultural practices. 
As further away we look in time, the harder the task. The scarcity of written sources, the need of a high command of ancient languages or the loss of the knowledge of ancient musical notations are just a few of the challenges that researchers must face in the search of the soundscapes of the past.

Part of the charm and the practical application of this discipline is the possibility to reenact forgotten music that has not been heard for centuries by recovering the aesthetics and performing styles lost to us. Most of them are not part of the modern performers' training. This is the work of 'early musicians' who apply what is known as historical informed performance. As in the analogous redshift approach in astronomy, the success of this activity is not free of caveats and is subject to ongoing debate \cite{Butt2002}. However, most researchers and practical early musicians acknowledge this is never-ending endeavour: we might never be able to know exactly how music at a particular time and culture sounded, but we can asymptotically approach the original source increasing the amount of evidence with continuous research and experimentation.

Although this started as a discipline focusing on Baroque, Renaissance and European Medieval music, these methods are being extended to other periods and cultures. Chinese \cite{Chen1989}, Japanese \cite{Nelson2018}, Korean \cite{Song1973} or Indian \cite{Widdess1979} early music, ancient music from Greece \cite{Mathiesen1981}, Rome \cite{Landels2000} or even Mesopotamia \cite{Dumbrill2017}, are also subjects of study (creating in the process new methodologies and sub-disciplines, \cite{Ibison1996,Greaney2020}) that allow to unveil the complex and multifaceted musics of the early and ancient world. This endeavour could be considered part of a wider discipline, cultural (historical) musicology, for which approaches from anthropology, sociology or philosophy of music are needed in combination with acoustics, music cognition, or psychology of music.

 A compelling illustration of this cross-cultural research is the modern revival of Tang dynasty (7th–10th century) music in China, which draws on archaeological discoveries, rich written records and treatises, notated music, and iconography. Musicologists and performers have reconstructed court music practices using replicas of period instruments and historical notation systems, offering new insights into the aesthetics and functions of early Chinese music \cite{Chen1991}. Closely linked is the study of \textit{gagaku}, the Japanese court music tradition, which preserves—among other elements—repertoires and instruments of Tang-dynasty Chinese origin, known as \textit{t\={o}gaku}. By comparing notational density, historical descriptions, and cross-cultural transmission timelines, some scholars have argued that these pieces were originally performed at significantly faster \textit{tempi} than in contemporary \textit{gagaku} performances \cite{Nelson2008}. The \textit{t\={o}gaku} repertoire, which largely derives from Chinese Tang secular banquet music rather than ceremonial court rites, is a distinction crucial for understanding both its original context and its later transformation within the Japanese court, where it was formalized for ritual and aesthetic purposes.
 
As in the analogy of astronomy, early music and cultural historical musicology provides listeners with snapshots of other temporal and spatial coordinates. In a way, is like an ethnomusicological time machine.

This 'otherness feeling' can be also experienced when listening to musical works within our own culture from the distant past. As we go further back in time, the practices, aesthetics, and mindset of our own seven-hundred-years-ago ancestors might be as different as those of other present-day societies. Thus, early music presents material outside our own experience and culture. Investigating old, long-forgotten repertoires can serve as a way to bridge connections with worlds that are vastly dissimilar to our own, thereby prompting us to challenge our preconceptions about the nature of music, its function, and its aesthetics.

Besides learning to play forgotten instruments, performing music from early times is a practice that requires the 'unlearning' of many of the practices, techniques and automatic sensibilities created after training in modern (classical) music at the conservatory. For instance, the language of modes in medieval European music is quite different from the tonal system learnt at standard courses, even if it sounds tonal to the modern ears. Modern musicians are accustomed to perceiving melodies within functional harmony, which makes it challenging to shift to hearing modal melodies in an unharmonized context. This shift in aural perspective is difficult because the habit of relating melodies to chords can interfere with our ability to hear a melody in relation to the 'final' of a mode without mentally hearing the harmony \cite{Mariani2017}. This may lead to incorrect evaluation of the mode, the selection of a proper drone (if any) or the modal affect (the semiotics and psychological mood linked to a mode). 

The challenge of detaching oneself from the modern music framework involves understanding, internalizing and appreciating a different conception of the performance of the music, breaking from the aesthetics canons of the present time. Period instruments have a distinct quality sound different from the modern 'classical' ones. In contrast to the metal strings employed in modern string instruments, early instruments used other materials available at that time, like gut strings in Europe or silk strings in China. The size, type of wood, shape, are different, providing a new whole auditory experience. There are important differences between the soundscape of a Baroque prelude played in a period harpsichord to the rendition of the same piece performed on a modern piano. 

This change in the aesthetic paradigm is even more noticeable in the use of the voice. The current modern opera vocal style with strong continuous vibrato, long legato or technique to project the voice might not have been in play in earlier times or at least, not in the same degree.
A clear example is the case of the French musicologist Marcel P\'er\`es. His approach to Gregorian chant is very different from the canonical and widely used Solesmes style of singing defined around the beginning of the 20th century. As P\'er\`es puts it, the Solesmes restoration is very much Romantic in taste, with long phrasing and lack of ornamentation \cite{Sherman2003}, quite different from the proposed by the French musicologist and his Ensemble Organum \cite{Lacavalerie2002}.

This evidence implies that although there is a historical link for any particular present-day musical culture with ground foundations in the past, that does not mean that modern listeners (including modern musicians) of that culture share the musical background, aesthetic appraisal, narrative or, in other words, how the music is conceived and heard, of their musical peers of the past \cite{Wieczorkowska2010,Haynes2007}.

By reconstructing the soundscapes of the past, historical musicology not only enriches our understanding of musical evolution but also challenges modern assumptions about musical universality. This historical perspective underscores the importance of cultural context in shaping auditory aesthetics, a theme that will be further explored in the following sections.

While historical musicology provides valuable insights into the evolution of musical practices within specific cultures, it is equally important to examine how these practices vary across different cultural contexts. Ethnomusicology, as a discipline, extends this inquiry by exploring the diversity of musical traditions and their underlying cultural frameworks. By comparing musical systems across time and space, we can better understand the extent to which musical principles are shaped by cultural context rather than universal laws. This cross-cultural perspective not only challenges the notion of musical universality but also highlights the rich tapestry of human musical expression. In the following section, we turn to ethnomusicology to explore how cultural diversity influences musical aesthetics and perception.

\subsection{Ethnomusicology: Challenging the Myth of Musical Universals}

The notion of universal musical principles has been debated historically and remains relevant in contemporary discussions on the role of culture in shaping human experience. In the Western philosophical tradition, the roots of this debate can be traced back to ancient philosophers such as Plato and Aristotle, who offered contrasting views on the nature of knowledge and how it is acquired. Pythagoras also contributed to this discourse by proposing a mathematical basis for musical tuning, which suggests the existence of universal principles underlying musical scales. Similar observations were made previously and independently (with its own characteristics) in other ancient civilizations such as China \cite{Falkenhausen1993} and India \cite{Rowell1992}, further highlighting the ongoing interest in exploring the idea of musical universals. The Western concept of \textit{musica universalis}, which later came to be known in Europe as \textit{musica spherarum} or 'music of the spheres', was tied to the geometrical proportions of celestial bodies and the vibrations they were believed to produce and therefore, those pure proportions should account for the beauty and harmony of music.

In the early discussions on musical universals, a significant focus was placed on pitch structures, specifically those related to scales and harmony. In the 19th century, Hermann von Helmholtz expanded upon Pythagoras' concepts of simple integer ratios. Based on physics, physiology, and psychoacoustics, he laid the foundations of a scientific theory of music perception \cite{Helmholtz1863}. His theory of consonance based on the harmonic overtone series continues to be the base of many modern studies. 
However, it is worth noting that he thought a mistake to make his theory of consonance (``a mere summary of observable facts'') the essential foundation of the theory of music, as he considered melody the essential basis of music \cite{Steege2012}. One of his main central ideas is that the modern tonality system was developed from a \textit{freely chosen} principle of style, and thus, other systems were developed from other different principles. For him, scales, modes and harmonic are a consequence of aesthetic principles and are not a consequence of natural laws.

By the last quarter of the 19th century, a number of studies appeared questioning the paradigm of music universals. Although in another very different context, music critic George Bernard Shaw ironically cited the famous Longfellow clich\'e in a 1890 review: ``Though music be a universal language, it is spoken with all sorts of accents'' \cite{Shaw1977}. The work of Alexander Ellis \cite{Ellis1885} was one of the first significant challenges to the idea of musical universals. Ellis developed a precise measuring system (the famous ``cents'') and used it to measure the tuning systems of various instruments from different cultures across Eurasia. He found that many intervals did not match Western tuning systems. Ellis argued that these findings posed a fundamental challenge to universal theories of pitch structure: ``the Musical Scale is not one, not natural, not even founded necessarily on the laws of the constitution of musical sound... but very artificial, and very capricious''. 

The field took another important momentum with the studies and seminal works around the beginning of the second half of the 20th century that in addition to the sound dimension (.e.g., scales, pitches), included the need to understand, as a three-part model, the ``conceptualization about music'' and ``behavior'' in relation to music \cite{Merriam1964}, that is, the role of music in culture. As there are many languages and dialects in the world, almost such diversity was found in music. As dialects, some musics could share characteristics that make them familiar across particular societies, like in the case of some languages within the Indo-European branch. However, others could be mutually unintelligible, as languages that do not share the same common root. Notwithstanding, within the thee-part analysis, it is not enough to share sound similarities as they might not express the same meanings (like false cognates, to continue with the language analogy). The concept of a major scale in classical European tonal music theory might not be understood or have the same cultural baggage in all musics \cite{Dean2022}. Thus, a musician versed in another 'language' might choose different structural patterns than those originally intended by the composer (e.g., example of the medieval modes in section \ref{sec:hm}). 

Cross-cultural studies show that there is no single universal hierarchy of the senses \cite{Levinson2018} and thus, culture is a important factor (i.e. nature vs nurture) that permeates human activity and the way we see, talk about and describe the world. In the case of ethnomusicology, an increasing number of works in the last decades have added valuable insights into the different modulations of music aesthetics. Studies in different regions of the world \cite{Messner1981,Ambrazevicius2017} show that the preference for consonance over dissonance is not universal. That preference is probably ``determined by exposure to musical harmony, suggesting that culture has a dominant role in shaping aesthetic responses to music'' \cite{McDermott2016}. The results of a more recent study \cite{Prete2020} reinforce the idea that the perception of musical pleasantness is not a shared experience across all individuals, but rather varies greatly depending on personal habits and cultural influences.

Over the past years, there has been a resurgence of interest in the psychology of music, leading to a revitalization of research on universals in this field. With an empirical approach, researchers aim to uncover the biological and cultural factors that contribute to the observed variability in music, and to determine the extent to which various aspects of music are present across different cultures. Gathering all this evidence, 
a list of 70 putative statistical universals grouped into several types was proposed, encompassing not only sound-related aspects (e.g., rhythm, melodic structure, texture, and pitch) but also contextual and behavioral dimensions \cite{Brown2013}. Some of them are quite broad and general, like the use of verbal text in vocal music or the use of music in religious or ritual contexts. Others are more related to physiological factors, like the length of vocal or aerophone instrumental phrases, that might be explained by physical constraints (lung capacity). However, most of them are very basic and when tested, many are indeed statistical universals but not certainly absolute musical universals \cite{Savage2015}. While some of these statistical universals may be constrained by human biology, they do not say anything on how these factors are organized in the sound discourse: cultures can combine and develop these set of basic universals in their unique ways to create music that holds distinct meanings for different listeners. Some simple, very basic musical emotions could be in some way shared, however complex emotions and above all, music \textit{and} its context are more idosyncratic and can be experienced differently. For instance, music associated with the phenomenon of death (sadness) in one culture is not necessarily understood in the same way in another musical culture \cite{Tagg1993}. What is clear is that music is universal in the sense that all known cultures display some kind of activity we can identify as music.

\subsection{Tuning Systems and Temperaments: The Cultural Fabric of Sound}

Although one of the statistical universals is the octave equivalence and the use of discrete pitches, in particular scales of seven of fewer pitches per octave, the actual separation of those discrete pitches is by no means universal \cite{Brown2013}. The harmony-of-the-spheres adepts would find surprising that the ``perfect'' intervals are not cosmic neither universal. 

Given the more o less general trend of the equivalence of the octave, this basic acoustic `distance' has been used in many cultures as a reference canvas to define a set of intervals for dividing it. Although the term `tuning' is usually understood in a broad sense as the fine-tuning of a musical instrument for its use, strictly speaking, the term should be used when intervals can be expressed as the ratio of two integers. As in astronomy, where the solar year cannot be filled with an integer number of lunar months, in music the octave cannot be circled with an integer number of just intervals. Specifically, it is not possible to tune octaves, just fifths and just thirds at the same time, as radical numbers are needed to express the ratios of some or all of its intervals. This modification of a tuning is called a temperament.

\subsubsection{Historical Development of Temperaments}

Along history, do\-zens of temperaments have been developed to deal with this issue. Only in Europe, more than 180 systems can be accounted since the Renaissance \cite{Barbour1951}. 
The relationships between intervals within a musical scale create a unique `flavor' or aesthetic character in the soundscape of a piece. Historical informed performance therefore strives to uncover the range of temperaments that were used during a specific period, in order to recreate the musical taste and style of the composers and musicians of that era.

These different ratios would elicit a very specific flavour when performing a European Baroque piece on a harpsichord written on a particular key  \cite{Young1991}. However, this fact is often forgotten by modern musicians that play Bach's works on a piano tuned in twelve-tone equal temperament (12-TET). To this date, there is still a common error to believe that Bach wrote the \textit{Well-Tempered Clavier} to promote equal temperament. While there is ongoing debate regarding the precise type of well \textit{unequal} temperament used by Bach, it is clear that it was not a ``modern'' 12-TET \cite{Rasch2011}, despite it was known during that time. 

When performed on European period keyboard instruments, unequally tempered scales embody a characteristic atmosphere, leading certain keys to be more fitting than others for conveying specific emotions through music, according to subjective associations of the composers of that time. As we saw before, the context is cultural dependent (i.e., association of a key with a particular affect), but the characteristic sounding difference, the acoustic fingerprint of each key on well-\textit{unequal} temperaments, is real. One may lead to believe this is a technicality of a very distant past, but studies show that these ideas and the aesthetic use of temperaments (different from the 12-TET) were at play well deep into the 19th century \cite{Steblin2002,Duguay2016}.

\subsubsection{Global Perspectives on Tuning}

Although the solution to the 12-tone equal temperament was first mathematically derived in China by Zhu Zaiyu in the second half of the 16th century \cite{Xu2021}, it was not used in common day music outside the complex cosmo-politics of the Court. As we have seen, this system was known later to musicians like Bach, but it was not widely adopted  in Europe as (the only) standard until the beginning of the 20th century. From then, it has become the system `to rule them all', to the extent that there is a general believe that is the `proper' (and only existing) way of tuning. However, as we have discussed, music of the past created in the framework of different interval relations gets `flattened' when performed within this system. This effect is even more relevant when playing music from other cultures using very different temperaments. We have seen that one statistical universal is that scales usually have seven of fewer pitches per octave. This does not imply that the relations between those notes are the same. The different scales that Ellis described  back at the end of the 19th century were almost, using an analogy with Kuhn's theory of science \cite{Kuhn1962}, `incommensurable' in the sense that there was no possible translation between their tuning/temperament systems. This tension is exemplified by Colin McPhee’s and Benjamin Britten’s transcriptions of Balinese gamelan, where the non-Western tunings of \textit{sl\'endro} and \textit{pelog} were inevitably reshaped by the constraints of Western notation and the 12-TET piano \cite{Church2021,Cooke1998}. Let's take for example the \textit{sl\'endro} scale used in Javanese Gamelan, a five-pitch equidistant tuning system (5-TET) \cite{Spiller2004}. The intervals of the \textit{sl\'endro} cannot be played on a 12-TET piano. It is rather common in Western music schools to teach the pentatonic scale and its different modes. However, they are usually exemplified on a piano, giving the false impression that all pentatonic scales (regardless of the mode or starting point of the scale) sound the same due to the `flattening' effect of a 12-TET piano. A person or even musician trained only in 12-TET will find difficult to sing in tune a Javanese piece set in \textit{sl\'endro} scale \cite{Perlman2004}. Needless to say the difficulty is presented also with other temperaments that divide the octave in many more intervals than twelve \cite{Tuma1996,Bor2010,Ederer2011}.

\subsubsection{Cultural Adaptivity in Sound Design}

For example, a sound design or sonification project using 12-TET may fail to resonate with audiences accustomed to other tuning systems, such as the Javanese \textit{sl\'endro} pentatonic scale. The \textit{sl\'endro} scale, with its unique intervals and cultural significance, produces a distinct soundscape that cannot be accurately replicated using 12-TET. A sound design based on \textit{sl\'endro} but rendered in 12-TET would not only sound different but could also be perceived as misleading or even disrespectful to listeners familiar with the original tuning. The use of a 12-TET pentatonic scale might give the impression that it is equivalent to the \textit{sl\'endro} scale, disregarding its cultural and aesthetic nuances. This highlights the need for culturally adaptive approaches in sound design that respect and incorporate aesthetic principles of diverse musical traditions.

Tuning systems and temperaments are intrinsic and fundamental building blocks of the aesthetic component of the music that work at force with other expressive variables. Different cultures and musical traditions have their own unique tuning systems (or even do not have discrete pitches), which can sound unfamiliar or even dissonant to those who are not accustomed to them. During the Victorian era in England, it was common for people to express difficulty in appreciating the unfamiliar India's classical music. The tuning of Indian \textit{raga} scales was criticized for being indicative of poor hearing because it did not match the tuning of the English audience \cite{Titon2008}. This is just an example of not trying to comprehend other's music within its own cultural context. Judging a music's tuning system by another standard often leads to the perception of it being `out of tune', even though it is actually perfectly in tune according to its own musical tradition.

\section{Discussion}

The idea that music is a ``universal language'' remains compelling to many people today, including musicians. This notion suggests that all types of music are inherently comprehensible to anyone, regardless of cultural background—a contrast often drawn with spoken languages, which tend to be mutually unintelligible. While this view reflects a genuine appreciation for music's emotional immediacy and cross-cultural resonance, it can also obscure important differences. Although some musical traditions may share common features (such as various forms of `classical' music within Europe), music and language are fundamentally distinct. Assuming universality can inadvertently imply that one musical tradition—its theories and aesthetics—has universal validity. This issue is especially salient in applied domains such as sound design and sonification, where auditory experiences are shaped by cultural context. A more inclusive approach to sound acknowledges the diversity of listening frameworks and interpretive practices across cultures, enriching our understanding of how sound communicates meaning.

\subsection{Cultural Bias in Music Theory and Technology}

A widely recognized example of cultural bias in music scholarship is the frequent emphasis on Western music within many standard music theory and history texts. Titles such as \textit{Fundamentals of Music Theory} or \textit{A History of Music}, often focus almost exclusively on Western musical traditions, which can inadvertently present them as the primary or most authoritative lens through which to understand music. While this emphasis may reflect the historical formation of academic music studies in many institutions, it may unintentionally marginalize non-Western musical practices and suggest that Western tonal principles are universally applicable. This perspective can reinforce the perception that Western music theory constitutes a normative or comprehensive system. Recognizing this pattern offers an opportunity to broaden our understanding of music by engaging with a wider range of theoretical approaches and musical systems, thereby enriching the field and inviting more inclusive perspectives.

Recent scholarship has highlighted the structural and institutional factors that perpetuate this bias. For instance, the concept of a ``white racial frame'' has been used to describe how unconscious biases influence the development of music theory, from the selection of repertoire to the recognition of authoritative theorists \cite{Ewell2020}. While diversity initiatives have sought to address these issues, their impact has been limited, underscoring the need for a more inclusive approach that incorporates non-Western perspectives \cite{Ewell2020}. This bias extends beyond theory and history into the realm of music technology, where Western-centric algorithms and standards (e.g., MIDI) may not adequately accommodate other musical aesthetics \cite{Moussa2002, Sarkar2012, Thompson2014} or influence the interpretation of research findings in the analysis of music \cite{Cornelis2010,Panteli2018}. As a result, culture-specific methodologies are often required for the analysis and representation of non-Western music \cite{Grupe2008,Serra2017}.

This bias also shapes music technology infrastructures, prompting scrutiny of how tools, classification systems, and software encode cultural assumptions. For example, George E. Lewis contrasts Afrological and Eurological approaches to human-computer improvisation, where African-American improvisational aesthetics—centered on complexity, collectivity, and real-time negotiation—are often marginalized or reframed through Eurocentric lenses that emphasize minimalism, hierarchy, and depersonalized sound \cite{Lewis1996,Lewis2000}. Similarly, a bibliometric analysis reveals racialized exclusions in musicological databases, where electroacoustic music is framed as ``white'' and ``academic'', while Black electronic artists are relegated to ``popular'' genres despite employing analogous techniques such as sampling and studio manipulation \cite{Sofer2020}. At the level of software architecture, computer music languages such as Max/MSP, SuperCollider, and ChucK embed idiomatic patterns—structures and workflows that are more accessible or ``natural'' within a given language—which shape compositional and performance practices even when such systems are nominally open-ended \cite{Mcpherson2020}. These patterns reflect deeper, often unacknowledged assumptions within the design, suggesting that no software can be entirely culturally neutral. Uncovering these foundational biases—frequently mistaken for universal concepts—not only reveals the cultural underpinnings of such systems but may also, paradoxically, open up new creative directions \cite{Puckette2002}. 

Recent critiques also emphasize the dominance of Eurocentric conventions in sound design, highlighting how interpretive cues, such as associations with water or technological sounds, are shaped by Western media and cinematic tropes. These cultural references limit the expressive potential of sound by privileging dramatized templates over diverse sonic logics. A pluriversal approach to sound design advocates for community-led strategies that embrace multiple aesthetic frameworks, fostering relational modes of meaning-making \cite{Hug2024}. Taken together, these perspectives reveal how both institutional frameworks and software infrastructures encode cultural and aesthetic ideologies, underscoring the need for critical engagement with the tools and categories that define contemporary sound practices.

This cultural encoding is particularly evident in the design of Digital Audio Workstations (DAWs), which are shaped by cultural assumptions inherent in software development practices. These tools often privilege Western musical norms—such as 12-tone equal temperament, grid-based metric structures, and visually quantized timing—while offering limited support for non-Western tuning systems, flexible or unmetered rhythms, and expressive performance techniques. This is not simply a legacy of historical hierarchies, but a consequence of our own collective participation in a development culture that tends to prioritize technical convenience, commercial viability, and familiarity with Western (pop) idioms \cite{Pardue2022O}. The dominance of standards like MIDI and the emphasis on backward compatibility reinforce these constraints, even as newer protocols like MIDI Polyphonic Expression (MPE) provide opportunities for more inclusive design. Recognizing this dynamic invites us to reflect critically on the assumptions we bring into music technology and to actively consider how alternative perspectives and musical systems might reshape the tools we create and use.

Even when we recognize that absolute universals in music are elusive, it can still be tempting to turn to statistical universals—gene\-ral trends observed across cultures—as a means of identifying common ground. While this approach can yield valuable insights, it also carries important limitations. Statistical universals are often too broad to capture the nuanced interplay among musical elements such as rhythm, pitch, and timbre. The aesthetic experience and evaluation of music—and of sound more generally—emerge from complex interactions among these elements, which can differ significantly even within a single cultural context. For example, in sonification, the same input data can give rise to soundscapes that are perceived and appraised in mutually exclusive ways by different audiences \cite{Garcia-Benito2022}. Furthermore, cross-cultural studies must carefully consider the roles of historical contact and cultural transmission. Apparent universals may reflect not intrinsic patterns, but rather the effects of inheritance, the adoption of practices \cite{Naroll1961}, or the broader legacies of colonialism and globalization. The global reach of Western music, for instance, is closely tied to these historical processes—including dissemination through colonial expansion, mass media, and the commercial music industry \cite{Kraus1989}. These dynamics present important considerations for research in music cognition and psychology, particularly when Western musical stimuli are used to examine emotional or perceptual responses across cultures \cite{putkinen_2024}. In addition, the musical databases employed in such research often lack diversity, drawing primarily from 20th-century Western music performed in modern stylistic frameworks.

\subsection{Tradition and Representation in a Globalized Musical Landscape}

Globalization has facilitated the international circulation of non-Western musical forms, often under the label of ``tradition''. However, this label can obscure the modern or hybrid nature of the music it describes. In some cases, new genres have been created or reshaped specifically for tourism or global consumption, yet are presented as ancient or culturally continuous. For example, the genre of \textit{kecak} in Bali—widely marketed as a traditional form—was in part developed at the beginning of the twentieth century through processes of cultural staging and selective reinvention  aimed at appealing to foreign audiences \cite{Johnson2002}. These cases demonstrate how tradition can be strategically constructed to respond to external demands, raising questions about historical authenticity, cultural agency, and the roles of mediation and adaptation. These concerns are compounded when Indigenous or community-specific knowledge systems are engaged through dominant academic or artistic frameworks. Some critiques have drawn attention to how settler-colonial modes of listening appropriate Indigenous sound practices, aestheticizing them without regard for their epistemological grounding or cultural protocols \cite{Robinson2020,Tan2008}. Even collaborative initiatives may perpetuate asymmetrical dynamics if they prioritize external musical expectations or interpretive authority. In such contexts, complex ceremonial or territorial practices risk being reduced to consumable sonic artifacts, while deeper systems of knowledge and sovereignty remain marginalized.

In light of these challenges, while globalization fosters valuable opportunities for intercultural exchange, it also necessitates greater attention to how music is framed, marketed, and understood across cultures. Cross-cultural collaborations, particularly in music, can provide a beautiful platform for sharing musical traditions and promoting mutual understanding. However, these interactions are not without complexities. Projects that integrate diverse styles often require compromises, such as adjustments to temperaments and aesthetics, which can change the original character of the music. Despite these adaptations, the resulting works are sometimes still labeled as ``traditional'', further complicating our understanding of these musical forms. This ambiguity can blur the lines between modern adaptations and historical practices, making it harder to accurately trace the evolution of musical aesthetics. For instance, the term ``traditional'' itself carries inherent ambiguity—how old must a piece of music be to be considered traditional? Should it be 50, 100, or even 400 years old? This lack of consensus makes it difficult to conduct diachronic studies or draw clear conclusions about historical aesthetics. As a result, studies using ``traditional'' music may inadvertently suggest that modern listeners can understand ancient aesthetics, even when the historical authenticity of the performance or the ``oldness'' of the piece is uncertain. These challenges underscore the need for a nuanced approach that acknowledges the complexities of commercialization and representation of music in a globalized world. In sum, while the sharing of musical knowledge through collaboration is invaluable, it is crucial to acknowledge the power dynamics at play and ensure that adaptations do not obscure the deeper historical and cultural significance of the music.

\subsection{Implications for Sound Design and Sonification}

These examples illustrate the complexity of interpreting data in music research. Caution is warranted when drawing general conclusions, particularly in studies that focus on a single musical culture or rely on modern music to evaluate \textit{present}-day tastes. Similar limitations apply to sound design and sonification, where cultural context plays a crucial role in shaping perception and interpretation. It is essential to acknowledge the scope and boundaries of our work, recognizing that our findings may not be universally applicable.

Musical perception and appreciation are deeply shaped by cultural background, with listeners unconsciously assigning different weights to various musical parameters. For instance, rhythmic complexity, harmonic richness, and melodic contour may hold different levels of significance depending on the musical traditions a listener is familiar with. A musician trained in one tradition may initially find another tradition's aesthetic principles less engaging simply because they emphasize different musical aspects. Western classical music, for example, places a strong emphasis on harmonic development, while rhythm often follows a regular, hierarchical structure \cite{temperley_2000, pressing_2002}. In contrast, musical traditions such as West African drumming or Arabic maqam-based music foreground intricate rhythmic or melodic elements, which can make Western classical harmony seem less compelling to those accustomed to such styles. In Ghanian drumming, rhythmic perception is guided by a repeating, non-isochronous pattern, with emphasis often falling at the end of a cycle rather than at the downbeat, as in Western traditions \cite{locke_1982, pantaleoni_1985}. These structural differences highlight how rhythmic perception is not universal but culturally shaped.

Listeners’ engagement with music is further influenced by their learned perceptual frameworks. Individuals raised with the asymmetric meters of Balkan folk music can easily perceive and synchronize with these complex rhythms, whereas those unfamiliar with the tradition often struggle \cite{hannon_2005}. These examples illustrate how musical appreciation is contingent upon prior exposure: a listener accustomed to one tradition’s priorities—e.g., complex harmony versus complex rhythm—may find another’s aesthetic either captivating or unremarkable, depending on how its structural elements align with their learned expectations. This suggests that aesthetic judgments of music—whether one finds a piece engaging or monotonous—are not intrinsic to the music itself but are instead shaped by the listener’s cultural and cognitive predispositions. 

In the context of sonification, this insight underscores the importance of considering cultural diversity when designing auditory experiences. For instance, incorporating instruments with culturally specific timbres or exploring scales or different tuning systems could be an initial step toward creating more inclusive sonifications \cite{Yu2019,Buehler_koto_2019,Buehler_viral_2020}. However, deeper explorations that account for the interplay of rhythm, melody, sound aesthetics, and cultural context are needed to fully address the diversity of auditory experiences.

Another important factor is the presence of shared musical or psychoacoustic cues for emotional expression between two different cultures, which is often cited as evidence for the cross-cultural nature of psychophysical cues in music. However, using the language metaphor introduced earlier, we may find similarities between distant cultural traditions, just as some languages share common roots—such as those in the Indo-European family. Yet, this does not necessarily mean that the languages (or the music) are mutually understood. Moreover, it does not imply that other cultural pairs will exhibit similar coincidences or that these findings can be universally extended, as the fundamental cultural and aesthetic building blocks may differ significantly. Therefore, caution is needed when extrapolating results based on comparisons between only a few cultures.

This principle also applies to sound design. Implementations of alert signals, soundscapes, musifications, or sonifications are implicitly designed with an intended audience in mind, often sharing a cultural background with their creators. Similarly, evaluations of these designs or products tend to be conducted within a similarly focused population. While modern aesthetics—often influenced by Western traditions—are widely disseminated due to globalization, they do not necessarily encompass or reflect other aesthetic perspectives. Taken together, these observations underscore the importance of considering cultural context in both musical research and practical applications, as universal claims based on limited cultural comparisons risk overlooking the diversity of human musical experience.

These considerations extend beyond music and sonification to encompass broader practices in sound design. For instance, the concept of ``good acoustics'' in concert halls has been shaped and naturalized within Western classical music culture \cite{Eidsheim2015}. Acoustic qualities such as reverberation, clarity, and intimacy have been defined, measured, and idealized in relation to prominent concert venues. Though grounded in scientific, musical, and architectural discourse, these criteria are widely internalized by audiences through repeated cultural exposure, even if not consciously understood. The traditional concert setting—characterized by a large, static, frontal listening position—reinforces a two-dimensional spatial conception of sound. This listening framework has become so normalized that deviations from it are often perceived as acoustical our sound design deficiencies. This “figure of sound” \cite{Eidsheim2015}—a combination of psychoacoustic experience, visual framing, and cultural habit—limits how we conceive and value sound, music, and spatial relationships. The naturalization of Western concert hall acoustics—with ideals such as a two-second reverberation time and frontal spatialization—has likewise influenced sound design practices more broadly.

A common issue in sonification is determining the degree of intuitiveness of the end product for the target audience. As we have seen in the case of music, some cognitive responses, although rooted in biological factors, are modulated by cultural influences \cite{Bender2016}. This has significant implications for sonification techniques, such as parameter mapping, particularly in determining intuitive mappings for data. For instance, studies on spatial-temporal mappings reveal that cultural attitudes toward time influence cognitive processes. Mandarin speakers, for example, conceptualize the past as ahead and the future as behind, in contrast to English or Spanish speakers \cite{Gu2019}. These cross-cultural differences in spatial-temporal mappings highlight how cultural frameworks shape perception and interpretation.

Similarly, in a musical or soundscape-oriented context, cultural differences can lead to counterintuitive mappings for certain audiences. For example, Tuvan throat-singers use low-pitched, chest-resonant `mountain' \textit{kargyraa} to metaphorically represent great heights, while higher-pitched `steppe' \textit{kargyraa} is used for other contexts \cite{Levin2006}. Such mappings, while intuitive within Tuvan culture, might not translate effectively to other cultural contexts. This underscores the need for careful consideration of cultural diversity in sound design and sonification.

When evaluating the intuitiveness or aesthetic appeal of a sonification or sound design, it is crucial to consider the cultural and musical background of the sample group. Questions such as who designed the test, the diversity of the audience, and the musical preferences of the participants (e.g., preferred genres, language, etc.) are essential to address. Without this contextual information, evaluations risk being culturally biased or limited in their applicability. These considerations highlight the importance of designing sound-based applications with a nuanced understanding of cultural diversity, ensuring that they resonate effectively across different audiences.

Rather than abandoning the search for common ground altogether, the critique of universalism calls for a shift in how commonality is conceived. Instead of presuming shared structures or values, inclusive design can begin by acknowledging the plurality of listening modes and musical ontologies, treating them not as deviations from a norm but as valid in their own right. This reframing opens the possibility for design practices that do not flatten difference but instead cultivate spaces for negotiation and mutual intelligibility. One essential step is to broaden musical education—not to produce experts in every tradition, an impossible task, but to foster exposure to, and respect for, a wider array of musical and aesthetic experiences. This kind of openness enables creators and designers to work alongside individuals who are deeply knowledgeable—whether as cultural insiders or academically trained experts—who can inform and guide the creative process. As suggested in research on musical beliefs, even limited engagement with unfamiliar musical systems can challenge ingrained assumptions and expand one’s conceptual repertoire \cite{Walker1990}. This exposure is not merely enriching on an individual level; it lays the groundwork for more equitable design paradigms, where different sonic logics can be appreciated on their own terms.

\section{Conclusions}

Music and aesthetic appreciation are not innate but acquired through learning and exposure. When the first Jesuits arrived in China in the early 17th century, they initially struggled to comprehend the music of their Chinese colleagues. Yet through dedicated study and genuine interest, they came to understand, appreciate, and even enjoy the art of their peers \cite{Garcia-Benito2021,Garcia-Benito2024}. This historical example reminds us that engaging meaningfully with unfamiliar sonic worlds is both possible and enriching.

In his discussion, Huron emphasizes the importance of examining cross-cultural distinctions in music cognition, while also cautioning against the potential loss of cultural diversity and therefore, of cognitive diversity \cite{Huron2004,Huron2008}. Sound artists, engineers, composers, and scholars with expertise in or from diverse cultural and musical backgrounds should be encouraged to envision alternative designs and sonification approaches so we create a more inclusive aural representation of the universe. 

As the field of sound design and sonification continues to evolve, it is essential to embrace cultural diversity as a core principle, ensuring that our work reflects the full spectrum of human musical expression. By doing so, we can create auditory experiences that are not only more inclusive but also more meaningful and impactful for global audiences. Without such consideration, we risk falling into the trap illustrated by the Indian parable of the blind men and the elephant—each of us grasping only a fragment, while the greater whole remains elusive.

\section{Acknowledgments}

The author thanks the three anonymous referees for their helpful comments and suggestions, which improved the clarity and presentation of the manuscript. Financial support from the Severo Ochoa grant CEX2021-001131-S, funded by MCIN/AEI/10.13039/ 501100011033, and project PID2022-141755NB-I00 is also acknowledged.

\bibliography{refs_au2025}

\end{document}